\documentclass[prd,amsmath,amssymb]{revtex4}
\usepackage{graphicx}

\usepackage[normalem]{ulem}  
\usepackage[dvips]{color} 

\def\x{{\boldsymbol x}}
\def\k{{\boldsymbol k}}

\def\u{{\boldsymbol u}}
\def\v{{\boldsymbol v}}

\def\bGamma{{\boldsymbol \Gamma}}
\def\wI{{\widetilde I}}
\def\wPi{{\widetilde \Pi}}

\def\z{{\boldsymbol z}}
\def\x{{\boldsymbol x}}
\def\y{{\boldsymbol y}}
\def\k{{\boldsymbol k}}
\def\q{{\boldsymbol q}}
\def\p{{\boldsymbol p}}
\def\0{{\boldsymbol 0}}
\def\v{{\boldsymbol v}}

\begin{document}

\begin{flushright}
KEK-TH-1477\\
UT-Komaba-11-5
\end{flushright}

\vspace*{1cm}

\title{Critical exponents from two-particle irreducible 1/$N$ expansion}

\author{Yohei Saito\footnote{corresponding author, 
e-mail: {\tt yoheis@post.kek.jp} }}

\affiliation{Department of Physics, Faculty of Science, 
University of Tokyo,\\ 7-3-1 Hongo Bunkyo-ku Tokyo 113-0033, Japan}

\author{Hirotsugu Fujii}
\affiliation{Institute of Physics, University of Tokyo, Tokyo 153-8902, Japan}

\author{Kazunori Itakura}
\affiliation{KEK Theory Center, IPNS,
          High Energy Accelerator Research Organization (KEK) \\
           1-1 Oho, Tsukuba, Ibaraki, 305-0801, Japan}
\affiliation{Department of Particle and Nuclear Studies, 
Graduate University for Advanced Studies (SOKENDAI), 
1-1 Oho, Tsukuba, Ibaraki 305-0801, Japan}

\author{Osamu Morimatsu}%
\affiliation{KEK Theory Center, IPNS,
          High Energy Accelerator Research Organization (KEK) \\
           1-1 Oho, Tsukuba, Ibaraki, 305-0801, Japan}
\affiliation{Department of Particle and Nuclear Studies, 
Graduate University for\\ Advanced Studies (SOKENDAI), 
1-1 Oho, Tsukuba, Ibaraki 305-0801, Japan}
\affiliation{Department of Physics, Faculty of Science, University of Tokyo,\\ 7-3-1 
Hongo Bunkyo-ku Tokyo 113-0033, Japan\\}

\begin{abstract}
We calculate the critical exponent $\nu$ in 
the $1/N$ expansion of the two-particle-irreducible (2PI)
effective action for the $O(N)$ symmetric $\varphi ^4$ model 
in three spatial dimensions. 
The exponent $\nu$ controls the behavior of a two-point
function $\langle \varphi \varphi \rangle$ 
{\it near} the critical point $T\neq T_c$, but can be evaluated 
on the critical point $T=T_c$ by the use of 
the vertex function $\Gamma^{(2,1)}$. We derive a self-consistent 
equation 
for $\Gamma^{(2,1)}$ within the 2PI effective action, and solve it 
by iteration in the $1/N$ expansion.
At the next-to-leading order in the $1/N$ expansion, 
our result turns out to improve those obtained 
in the standard one-particle-irreducible calculation.

\end{abstract}

\maketitle

\newpage
\section{Introduction}

Understanding equilibrium and nonequilibrium phenomena associated 
with phase transitions has become 
more and more important in various fields in physics,
such as early-time universe,
ultra-relativistic heavy-ion collisions, 
ultra-cold atoms, and so on \cite{CalHu,Boyanovsky:2006bf}. 
In a second-order phase transition,
characteristic long-range fluctuations appear in the order 
parameter field, and for a quantitative study of 
static and dynamic critical phenomena \cite{Hohenberg,Mazenko}, one needs a field
theoretical method which can describe both static and dynamical
processes involving strong fluctuations.
Emergence of long-range fluctuations makes naive perturbation theory break
down and requires some sort of resummation, such as the method of the
renormalization group or the two-particle-irreducible (2PI) effective
action \cite{LW,CJT}.

Recently, the method of the 2PI effective action has received much
attention since it can be applied to the phenomena 
in and out of equilibrium on an equal footing \cite{Berges,ABC,JSD}. 
In this method, all the self-energy contributions for the two-point
correlation function are first summed up and then the perturbative
expansion is carried out in terms of the full two-point correlation
function. 
This is in contrast to the standard method of the
one-particle-irreducible (1PI) effective action in which the
perturbative expansion of the diagrams is done
in terms of the free two-point correlation function. 
The method of the 2PI effective action systematically resums higher
order terms in powers of coupling constants,
so that it is expected to take into account efficiently
the large fluctuations near the critical point.

In the present paper we restrict ourselves to static critical
phenomena and leave dynamic critical phenomena for future study. 
As is well known,
the most prominent feature of static critical phenomena is universality.
In other words, several critical exponents which characterize the singularities
in the vicinity of the critical point are solely determined
by symmetry of the system, irrespective of microscopic details. 
In fact, only two of them are independent,
and we take $\eta$ and $\nu$ to be studied in this paper.
They can be read off from the two-point correlation function 
$G(\x,\0) = \langle \varphi(\x)\varphi(\0)\rangle$ of the order parameter field
as
\begin{eqnarray}
  && G(\x,\0) \sim \frac{1}{|\x|^{d-2+\eta }} \qquad  (T=T_c)\, , 
  \label{eq:eta-x}
\\
  && G(\x,\0)  \sim {\rm e}^{-|\x|/\xi} \, ,\quad 
                     \xi \sim |T-T_c|^{-\nu} \qquad (T > T_c)\, ,
  \label{eq:xi}
\end{eqnarray}
where $d$ is the number of space dimensions (now $d=3$),  
$\xi $ is the correlation length, and $T_c$ is the critical temperature. 
Namely, $\eta$ and $\nu$ govern the behavior of two-point functions 
{\it on} and {\it off} 
the critical point, respectively.
In the momentum space,
the Fourier transform of Eq.~$(\ref{eq:eta-x})$ gives the
scaling form $\widetilde G(\k) \sim {|\k|^{\eta -2 }}$.

Recently, Alford, Berges, and Cheyne employed the $1/N$ expansion of 
the 2PI effective action to compute the 
exponent $\eta$ of an $O(N)$-symmetric $\varphi^4$ theory in 
three dimensions \cite{ABC}. 
They solved the 2PI Schwinger-Dyson equation (Kadanoff-Baym equation) \cite{KB paper,Baym,KB text}
at the critical point, substituting the scaling form to $\widetilde G(\k)$. 
It was shown that at the next-to-leading order (NLO) in the $1/N$ expansion 
this approach remedies the spurious divergence of $\eta$
at small $N$, which is seen in the 1PI $1/N$ expansion, 
and leads to an improved estimate already for moderate values of $N$.
This success strongly motivated us to compute another exponent 
$\nu$ within the 2PI effective action. The exponent $\nu$ is 
also associated with the critical behavior of the two-point functions.

It is not straightforward, however, to apply this method
to the evaluation of $\nu$.
Given two nonzero parameters, $p$ and $T-T_c$, one needs to fix the
form of $\widetilde G(\p; T)$ in the scaling region,
which introduces a technical complication to the problem.
Fortunately, we notice that the exponent $\nu$ 
can be determined from the {\it three-point vertex function} 
$\Gamma^{(2,1)}(\x,\y;\z)\sim 
\langle \varphi(\x)\varphi(\y)\varphi^2(\z)\rangle $
with two elementary fields, $\varphi$,  and one composite operator, 
$\varphi^2$ {\it evaluated at} $T=T_c$ \cite{Amit}. 
In fact, its Fourier transform 
(see Eq.~(\ref{eq:Gamma21def}) for definition)
{\it at the critical point}
behaves as
\begin{eqnarray}
  \widetilde \Gamma ^{(2,1)}\left(\frac\k{2},\frac\k{2};\k\right) 
  \sim |\k|^{2-\eta-1/\nu } \qquad (T=T_c)\, .
  \label{eq:Gamma21}
\end{eqnarray}
Therefore, if one finds the equation 
for $\Gamma^{(2,1)}$ in the 2PI formalism,
one should be able to compute $\nu$ at the critical point, similarly
to the case of $\eta$.

In this paper we develop the 2PI formalism for 
the three-point vertex function $\Gamma^{(2,1)}$, and apply the 
$1/N$ expansion to compute the exponent $\nu$.
We calculate 
$\Gamma ^{(2,1)}$ at 
the next-to-leading order in the 2PI $1/N$ expansion 
assuming the scaling form of the correlation function at the 
critical point.
We then extract the exponent $\nu$ according to Eq.~(\ref{eq:Gamma21})
and examine whether an improvement similar to the calculation of
$\eta$ is achieved.


Computation of the critical exponents has been challenged
since 70's,
in  the $\epsilon$-expansion \cite{Wilson} and $1/N$-expansion \cite{Ma,Ma2,Ma3,AbeHikami} 
appoarches in the 1PI effective action formalism. Furthermore,
the four-particle-irreducible (4PI) effective action has also been
applied in \cite{Vasil'ev}
to get the higher order terms in the $1/N$ expansion.
These methods are utilized to obtain a strict $1/N$ expansion series
eventually. In contrast, our motivation here is 
to examine a possible improvement due to 
the self-consistent approximation provided in the 2PI formalism.

This paper is organized as follows. In the next section, 
we first define our model and then explain how $\Gamma^{(2,1)}$ is 
related to the critical exponent $\nu$.
The formalism with the 2PI effective action is introduced
in the third section, where 
we also derive a self-consistent equation for $\Gamma ^{(2,1)}$. 
Then, in the fourth section, we calculate $\nu $ in the 
2PI effective action and compare it with the 1PI result,
where some complications in the calculations are deferred to Appendix. 
The final section is devoted to a summary of our results and 
discussions.

\section{Three-point vertex function $\Gamma^{(2,1)}$ and  
critical exponent $\nu$}

We consider an $O(N)$ symmetric $\varphi ^4$ model 
($\varphi_a=\varphi _1,\cdots ,\varphi _N$) 
in the three-dimensional Euclidean space. 
The action is given by 
 \begin{eqnarray}
  S[\varphi ]
   =\int d^3 x 
    \left[\frac{1}{2}\partial_i \varphi_a(\x) \partial_i \varphi_a(\x)
    +\frac{\lambda }{4! N}(\varphi _a(\x)\varphi _a(\x))^2
    +\frac{1}{2}\, t\, \varphi _a(\x)\varphi _a(\x) \right] ,
    \label{eq:action}
 \end{eqnarray}
where $t$ can be identified as either the mass squared or $T-T_c$, 
up to renormalization. Roughly speaking, the ground state is in 
a symmetric (broken) phase when $t>0$ ($<0$), and the transition 
at $t=0$ is of the second order. 
Although the exponents are symmetrical about $t=0$, 
we compute the critical exponents by approaching the critical point
from the symmetric phase ($t>0$)
because the vanishing expectation value
$\phi = \langle \varphi \rangle=0$ makes the computation 
technically less involved.

The two-point correlation function, $G_{ab}$, 
and its Fourier transform, $\widetilde G_{ab}$,  are defined as 
\begin{eqnarray}
 G_{ab}(\x,\y) 
  &\equiv& \left< \varphi_a (\x) \varphi_b (\y) \right> \nonumber \\
  &=& \int \frac{d^3k}{(2\pi)^3}\, {\rm e}^{-i\k \cdot (\x-\y)} 
      \, \widetilde G_{ab} (\k) \, ,
\label{eq:Gdef}
\end{eqnarray}
where translational invariance in the equilibrium state is assumed.
Being in the symmetric phase, we treat $G_{ab}$ as diagonal and we write
$G_{ab}=G \delta _{ab}$ unless otherwise stated. 
Similarly the three-point vertex function with two elementary fields
and one composite field,  $\Gamma ^{(2,1)}$, and its Fourier transform
are defined as 
\begin{eqnarray}
  \Gamma_{ab} ^{(2,1)} (\x,\y;\z) &\equiv & \int d^3x_1 d^3y_1 
\left< 
    \varphi_{a'} (\x_1) \varphi_{b'} (\y_1)\;  \tfrac{1}{2}\varphi_c^2(\z)
\right> 
\,     G_{aa'}^{-1}(\x,\x_1) \, G_{bb'}^{-1}(\y,\y_1)  \nonumber \\
  &=& \int \frac{d^3k}{(2\pi)^3} \frac{d^3p}{(2\pi)^3}\, 
   {\rm e}^{i\k \cdot (\x-\y)}\, 
   {\rm e}^{-i\p \cdot (\z-\y)}\, 
   \widetilde \Gamma_{ab} ^{(2,1)} (\k,\p-\k;\p)
\; ,
  \label{eq:Gamma21def}
 \end{eqnarray}
where summation over the indices $a',b'$ and $c$ is implied 
and we have assumed translational invariance of an equilibrium state.

Notice that there is a relationship between the 
two-point function $G$ and the three-point vertex function $\Gamma^{(2,1)}$. 
If one regards $t$ as the external 
field coupled to $\frac12 \varphi^2$, then the differentiation of $G$
with respect to $t(\z)$ gives \cite{Justin,Amit}
\begin{align}
\frac{\delta G_{ab}(\x,\y)}{\delta t(\z)}
= - \langle \varphi_a(\x)\varphi_b(\y)\tfrac12 \varphi_c^2(\z)\rangle \, .
\end{align}
Using 
$\delta G^{-1}/\delta t=-G^{-1} (\delta G/\delta t) G^{-1}$ which follows
from the identity $G^{-1} G=1$, one finds that the definition 
of $\Gamma^{(2,1)}$ yields 
\begin{equation}
\Gamma^{(2,1)}_{ab}(\x,\y;\z)=
\frac{\delta G_{ab}^{-1}(\x,\y)}{\delta t(\z)}\, .
\label{eq:GammaG}
\end{equation}
The corresponding equation holds in the momentum space. 
In particular, in the zero momentum limit, one has 
\begin{align}
  \widetilde \Gamma ^{(2,1)}(\0,\0;\0)
&=\frac{\partial \widetilde G^{-1}(\0)}{\partial t}
\; .
\end{align}

At the critical point, $t=0$, 
the exponent $\eta$ is determined from the low-momentum behavior of 
the two-point correlation function $\widetilde G(\k) \sim 
 |\k|^{-2+\eta}$, 
while the exponent $\nu$ can be obtained from 
$\widetilde \Gamma^{(2,1)}(\k,\p-\k;\p)$
as shown in Eq.~(\ref{eq:Gamma21}) \cite{Ma}.  
This can be explained with the help of the scaling hypothesis
applied to $\widetilde \Gamma^{(2,1)}(\k,\p-\k;\p)$.
Near the critical point,
the  susceptibility, $\chi$, behaves as
$ \widetilde G(\0) = \chi \sim t^{-\gamma}$
with the critical exponent $\gamma$, which  immediately 
implies that 
\begin{eqnarray}
  \widetilde \Gamma ^{(2,1)}(\0,\0;\0) \sim t^{\gamma-1}.
  \label{dif:t}
\end{eqnarray}
In the scaling hypothesis
we assume the existence of a function $f$ and a constant $y$ such that 
($k=|\k|$)
 \begin{eqnarray}
  \widetilde\Gamma ^{(2,1)}\left(\frac\k{2},\frac\k{2}; \k\right) \sim f(k\xi ) \xi ^y\, . 
  \label{fy}
 \end{eqnarray}  
When $t \ne 0$, the limit $k\to 0$ is regular and so is 
$f(0)$, which yields
 \begin{eqnarray}
  \widetilde \Gamma ^{(2,1)}(\0,\0;\0) 
 \sim f(0) \xi ^y \sim t^{-\nu y}\, , 
  \label{fy0}
 \end{eqnarray} 
where the use has been made of $\xi=t^{-\nu}$ (see Eq.~(\ref{eq:xi})).
Comparing Eqs.~(\ref{dif:t}) and (\ref{fy0}),
one finds 
 \begin{eqnarray}
  y=\frac{1-\gamma }{\nu }\, . 
  \label{y}
 \end{eqnarray}
As we approach the critical point $t \to 0$,
the correlation length $\xi$ diverges, 
while $\widetilde \Gamma ^{(2,1)}(\k/2,\k/2;\k)$ is still finite as 
long as $k$ is kept nonzero.
Therefore, we must have $f(k\xi ) \sim (k\xi)^{-y}$ 
to find the scaling form at the critical point:
 \begin{eqnarray}
  \widetilde \Gamma ^{(2,1)}\left(\frac\k{2},\frac\k{2};\k\right) 
  \sim k^{-y} \sim k^{\frac{\gamma -1}{\nu }}. 
  \label{xi and mu}
 \end{eqnarray}
This is equivalent to Eq.~(\ref{eq:Gamma21}) with the aid of
the scaling law $\gamma =\nu (2-\eta )$.
Diagramatically, it is shown as 
\begin{eqnarray}
 \widetilde \Gamma ^{(2,1)} \left(\frac\k{2},\frac\k{2};\k\right) 
= 
\; 
\raisebox{-5mm}{\includegraphics[width=1.8cm]{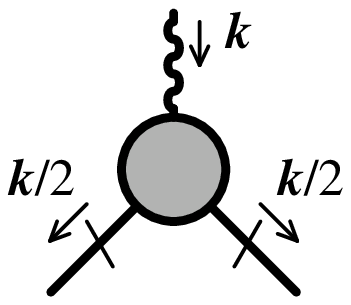}}
\;
\sim k^{2-\eta-1/\nu }\, ,
\label{eq:Gamma^(2,1)}
\end{eqnarray}
where a blob, a wiggly line and a simple line represent 
$\widetilde \Gamma ^{(2,1)}$, $\varphi ^2$ and $\widetilde G$, respectively. 
A slash on a simple line indicates the amputation.

\section{2PI effective action}
We give here a minimal review on the 2PI effective action, 
together  with the 1PI effective action for comparison.
The generating functional or the
partition function $Z[J]$ with an external field $J$ is
 \begin{eqnarray}
  Z[J]\equiv \int {\cal D}\varphi\, {\mathrm{exp}} 
\left[
-S[\varphi ]  +\int d^3x J_a(\x) \varphi _a(\x) 
\right]
\equiv {\rm e}^{-W[J]}\,
  , 
  \label{eq:Z[J]}
 \end{eqnarray} 
where $W[J]$ is the generating functional for the connected Green's
functions. The averaged field is given by
 \begin{eqnarray}
  \phi _a(\x) \equiv \left< \varphi _a (\x)\right> 
=\frac{\delta  W[J]}{\delta J_a (\x)}\, . 
 \end{eqnarray}
The 1PI effective action $\bGamma_{\rm 1PI}$ 
as a function of $\phi$ is obtained by the 
Legendre transformation of the generating functional $W[J]$, 
 \begin{eqnarray}
  \bGamma _{\rm 1PI}[\phi ]
\equiv
 W[J]-\int d^3xJ_a(\x) \frac{\delta W[J]}{\delta J_a (\x)}\, .
 \end{eqnarray}
Diagramatically, $\bGamma_{\rm 1PI}$ consists of the vacuum diagrams
written in terms of the lines representing the free two-point function $G_0(\phi)$ 
in the presence of the classical field, $\phi$. 
Each of 1PI diagrams does not split into two by cutting only one line.

The ground state is determined by the condition: 
$\delta  \bGamma_{\rm 1PI}[\phi]/\delta \phi_a(\x) = J_a(\x) = 0$,
which has a useful form for variational analysis.

In order to obtain the 2PI effective action, 
we introduce two external fields $J$ and $K$ and define the 
generating functional $Z[J,K]$ as
 \begin{eqnarray}
  Z[J,K]\equiv \int {\cal D}\varphi\, {\mathrm{exp}} 
  \left[-S[\varphi ] 
  +\int d^3x J_a(\x) \varphi _a(\x) 
  +\int d^3x d^3y \varphi _a(\x)K_{ab}(\x,\y)\varphi _b(\y)
  \right] 
  \equiv {\rm e}^{-W[J,K]} .
  \label{eq:Z[J,K]}
 \end{eqnarray}
Here the generating functional $W[J,K]$ is defined by the last equality.
The averaged field $\phi_a (\x)=\left< \varphi_a (\x)\right>$ and 
the full propagator (or the correlation function) 
$G_{ab}(\x,\y)= \left< \varphi_a (\x)\varphi_b (\y)\right> _{\rm connected}$
are respectively given by
 \begin{eqnarray}
  \frac{\delta W[J,K]}{\delta J_a (\x)}=\phi _a(\x), \qquad
  \frac{\delta W[J,K]}{\delta K_{ab} (\x,\y)}=
  \frac{1}{2} \left[G_{ab}(\x,\y) +\phi _a(\x) \phi _b(\y)\right]\, .
  \label{eq:W derivative}
 \end{eqnarray}
Performing the Legendre transformation of $W[J,K]$ with respect 
to $J$ and $K$, we obtain the 2PI effective action 
$\bGamma_{\rm 2PI}[\phi ,G]$ as a function of $\phi$ and $G$,
 \begin{eqnarray}
  \bGamma_{\rm 2PI}[\phi ,G]\equiv W[J,K]
  -\int d^3x\, J_a(\x) \frac{\delta W[J,K]}{\delta J_a (\x)}
  -\int d^3xd^3y\, K_{ba}(\y,\x) \frac{\delta W[J,K]}{\delta K_{ab}(\x,\y)} .
   \label{eq:double Legendre}
 \end{eqnarray}

One can explicitly extract the one-loop contributions from
the 2PI effective action
in the same manner as in the 1PI effective action (but now using the full
propagator), yielding the most general and useful 
form (for derivation, see Ref.~\cite{Berges}):
 \begin{eqnarray}
  {\bf \Gamma }_{\rm 2PI}[\phi ,G]
=
S[\phi ]+\frac{1}{2} \mathrm {Tr} \ln G^{-1}
+\frac{1}{2} \mathrm{Tr} G_0 ^{-1}G 
+\overline {\bf \Gamma }_2[\phi ,G]\, ,
  \label{eq:2PI}
 \end{eqnarray}
where Tr should be understood as integration over the space
  coordinates and summation over the field components.
The last term $\overline {\bf \Gamma }_2$  represents contributions from 
2PI vacuum diagrams 
in terms of the {\it full} propagator $G$, 
not of the free propagator $G_0$.

The ground state is determined by the stationary conditions with
respect to $\phi$ and $G$ at vanishing external fields $J=K=0$,
and turns out to be the same as in the 1PI effective action,
as it should.

\subsection{Self-consistent equation for $G$: Kadanoff-Baym equation}
Performing the functional derivative of Eq.~(\ref{eq:2PI}) 
with respect to $G$  and setting $K=0$, we obtain
$$
  0=-\frac{1}{2}G_{ab}^{-1}(\x,\y)+\frac{1}{2}G_{0,ab}^{-1}(\x,\y)
    +\frac{\delta \overline {\bf \Gamma} _2[\phi ,G]}{\delta G_{ba}(\y,\x)} .
$$
Comparing this with the Schwinger-Dyson equation, 
$G^{-1} = G_0^{-1} - \Sigma$ with the proper self-energy $\Sigma$,
we find that
 \begin{eqnarray}
  \Sigma _{ab}[\phi,G(\x,\y)]=
-2 \frac{\delta \overline {\bf \Gamma}_2[\phi ,G]}
        {\delta G_{ba}(\y,\x)}\, . 
  \label{eq:self-energy}
 \end{eqnarray}
Namely, the functional derivative of $\overline {\bf \Gamma}_2$ 
is identified with the proper self-energy, which must be 1PI, 
and therefore $\overline \bGamma_2$  is 2PI in terms of the full 
propagator $G$, as we mentioned above.
Thus, we arrive at a self-consistent equation for $G$, the
Kadanoff-Baym (KB) equation \cite{KB paper,Baym,KB text}:
 \begin{eqnarray}
  G_{ab}^{-1}(\x,\y)=G_{0,ab}^{-1}(\x,\y)-\Sigma _{ab}[\phi,G(\x,\y)]\, .
  \label{eq:KB}
 \end{eqnarray}
We remark here the following: if one eliminates $G$ in favor of $\phi$ from
$\bGamma_{\rm 2PI}[\phi,G]$ using Eq.~(\ref{eq:KB}) to obtain 
$\bGamma_{\rm 2PI}[\phi,G(\phi)]$ as the functional of $\phi$, 
one should formally recover the 1PI effective action 
$\bGamma_{\rm 1PI}[\phi]$, and therefore the ground states in 
both approaches must be the same. 
In practice, however, these effective actions are  different in
approximation level because resummation has been done in
$\bGamma_{\rm 2PI}[\phi, G(\phi)]$ \cite{Hees}.
Introduction of the full propagator $G$ satisfying the self-consistent
equation (\ref{eq:KB}) provides us of a way to reorganize the
expansion series in a perturbation theory.

\subsection{Self-consistent equation for $\Gamma ^{(2,1)}$}

Now one can derive the self-consistent equation for 
$\Gamma^{(2,1)}$ from the KB equation (\ref{eq:KB}) for $G$. 
By differentiating Eq.~(\ref{eq:KB}) with respect to $t(\x)$ 
and using the relation (\ref{eq:GammaG}), one obtains 
\begin{eqnarray}
\Gamma_{ab}^{(2,1)}(\x,\y ; \z) =
\Gamma_{0,ab}(\x,\y ; \z)  - 
\frac{\delta \Sigma _{ab}[G(\x,\y)]}{\delta t(\z)}
\, ,
\end{eqnarray}
where $\Gamma_{0,ab}(\x,\y;\z)=\delta(\x - \z)\delta(\y - \z)\delta_{ab}$.
Because we are in the symmetric phase $\phi=\langle \varphi \rangle =0$,
we have $G_{ab}=G \delta_{ab}$, $\Sigma_{ab}=\Sigma \delta_{ab}$, and
$\Gamma_{ab}^{(2,1)}=  \Gamma^{(2,1)} \delta_{ab}$, and thus we deal with 
the scalar functions without indices, hereafter.
Applying the chain rule for $\Sigma=\Sigma[\phi=0,G]$, 
we can rewrite the second term as 
 \begin{eqnarray}
\frac{\delta \Sigma(\x,\y)}{\delta  t(\z)} &=& 
\int d^3x_1 d^3y_1 
\frac{\delta G(\x_1,\y_1)}{\delta t(\z)} 
\frac{\delta \Sigma(\x,\y)}{\delta G(\x_1,\y_1)} 
\nonumber \\
&=& - 
\int d^3x_1 d^3y_1 d^3x' d^3y'
G(\x_1,\x')
 \frac{\delta G^{-1}(\x',\y')}{\delta t(\z)} 
G(\y',\y_1)
\frac{\delta \Sigma(\x,\y)}{\delta G(\x_1,\y_1)} 
\nonumber \\
&\equiv& - 
\int d^3x' d^3y'
D(\x,\y;\x',\y') \Gamma^{(2,1)}(\x',\y';\z)
\; ,
\label{eq:GG}
\end{eqnarray}
where we have used 
$\frac{\delta }{\delta t}G
   =-G\left(\frac{\delta }{\delta t}G^{-1}\right)G,$ 
similar to the one used for Eq.~(\ref{eq:GammaG}), 
and defined the kernel $D$ as
\begin{eqnarray}
D(\x,\y;\x',\y')
\equiv \int d^3x_1 d^3y_1 
G (\x_1,\x')
   \frac{\delta \Sigma (\x,\y)}{\delta G(\x_1,\y_1)} 
G(\y',\y_1) 
\; .
\label{eq:D4C}
\end{eqnarray} 
Thus, we write  the self-consistent equation for $\Gamma^{(2,1)}$ as
 \begin{eqnarray}
  \Gamma ^{(2,1)} (\x,\y;\z) 
  = \Gamma_{0} (\x,\y;\z)  + 
\int d^3x' d^3y' D(\x,\y;\x',\y') \Gamma^{(2,1)} (\x',\y';\z)
\; .
\label{eq:coord}
 \end{eqnarray}
In the momentum space, we have
 \begin{eqnarray}
  \widetilde \Gamma^{(2,1)} (\p,\q;\p+\q) 
  = 1  + \int \frac{d^3p'}{(2\pi)^3} \frac{d^3q'}{(2\pi)^3} 
\widetilde D(\p,\q;\p',\q') 
  \widetilde \Gamma^{(2,1)} (\p',\q';\p'+\q')
\; ,
\label{eq:mom}
 \end{eqnarray}
where the Fourier transform of the kernel is defined by 
 \begin{eqnarray}
 \widetilde D(\p,\q;\p',\q')
= \int d^3x d^3y d^3x' d^3y' 
    {\rm e}^{i(-\x\cdot \p -\y\cdot \q+\x'\cdot \p' +\y'\cdot \q')}
    D(\x,\y;\x',\y')\, .
 \end{eqnarray}
The kernel  $\widetilde D(\p,\q;\p',\q')$ contains the
momentum conservation condition $(2\pi)^3 \delta (\p+\q-\p'-\q')$. 
We evaluate the equation (\ref{eq:mom}) to calculate the vertex function
$\widetilde \Gamma^{(2.1)}$ with a given kernel $\widetilde D$ at the critical point $t=0$,
and determine the exponent $\nu$.

\section{Critical exponents from 2PI effective action}

As we explained in the previous sections, 
the exponents $\eta$ and $\nu$ are respectively associated with 
the two point function $G$ and the three-point vertex function 
$\Gamma^{(2,1)}$ at the critical point.
To determine the long distance behavior of these functions, 
one needs to solve the self-consistent 
equations ({\it i.e.}, Eqs.~(\ref{eq:KB}) and (\ref{eq:mom})),
both of which are derived from the 2PI effective action
$\overline {\bf \Gamma}_2$.

Let us first evaluate $\overline {\bf \Gamma}_2[G]$ (\ref{eq:2PI}) 
up to the NLO accuracy in the $1/N$ expansion. 
The LO and NLO contributions are respectively 
 \begin{eqnarray}
  \overline {\bf \Gamma} _2 ^{\rm LO}[G] &=&
\;
\raisebox{-2.5mm}{\includegraphics[width=1.5cm]{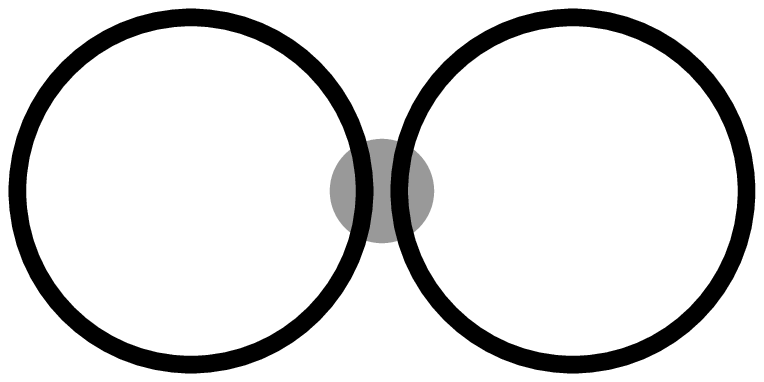}}
\;
 =\ -\frac{\lambda }{4!N} \int d^3x G_{aa}(\x,\x)G_{bb}(\x,\x)\, ,
  \label{eq:LO bubble} \\
\overline {\bf  \Gamma} _2 ^{\rm NLO}[G] &=&
\;
\raisebox{-7mm}{\includegraphics[width=1.5cm]{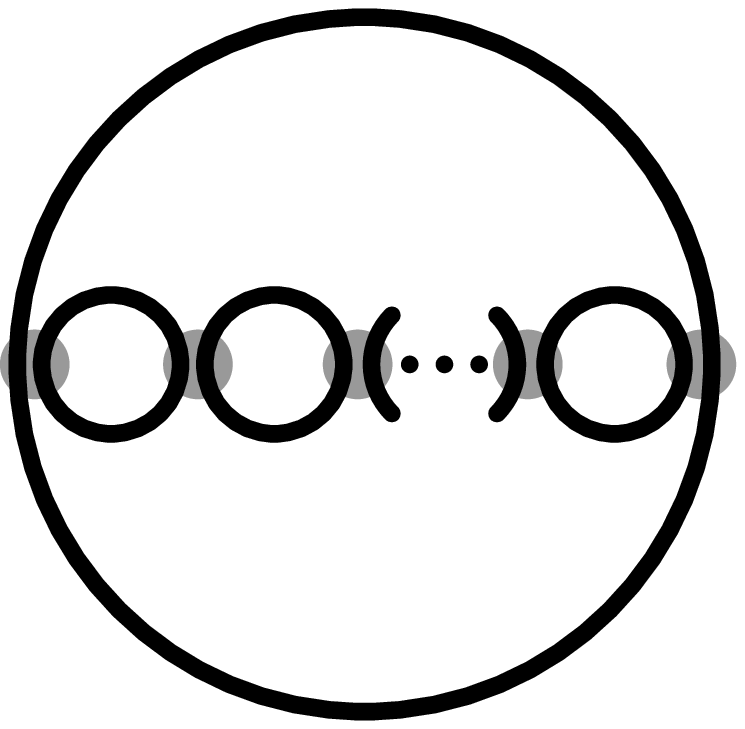}}
\;
  = \ \frac{1}{2} \mathrm{Tr} \ln \left[ \delta (\x-\y)
+\frac{\lambda}{6N} G_{ab}(\x,\y)G_{ab}(\x,\y) \right]\, , 
\end{eqnarray}
where a gray blob indicates a vertex $\lambda/N$ and a line corresponds
to $G_{ab}=G\delta_{ab}$. Summation over the repeated indices $a$,
$b$, etc.\ should be understood. Then a closed loop gives the number 
of the field components $N$, and thus the first diagram amounts to 
${\cal O}(N^2/N)={\cal O} (N)$, while the second ${\cal O}(N^0)$.
One obtains the self-energy at the NLO by cutting one
propagator in these diagrams, which yields in the momentum space 
 \begin{eqnarray}
  \widetilde \Sigma_{ab} (\p)&=&\ 
\;
\raisebox{-3mm}{\includegraphics[width=1cm]{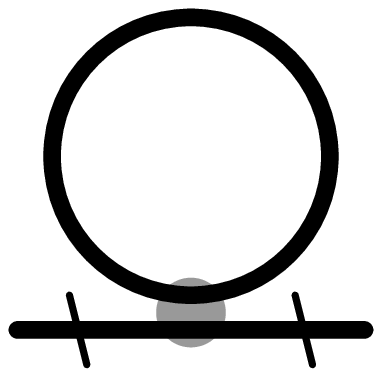}}
\;
+
\;
\raisebox{-3mm}{\includegraphics[width=1.8cm]{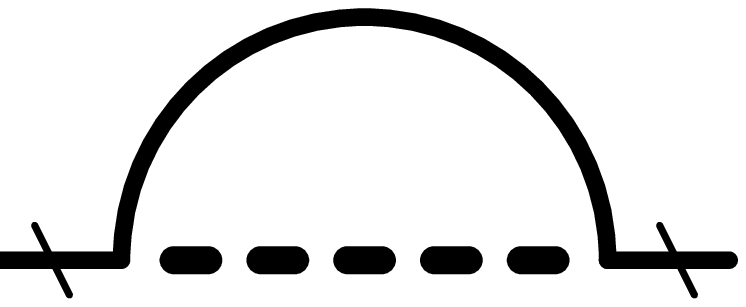}}
\;
 \nonumber \\
   &=& -\frac{\lambda }{6} \int \frac{d^3q}{(2\pi) ^3}
\widetilde G(\q)\delta_{ab}
-\int \frac{d^3q}{(2\pi)^3} 
\widetilde I(\q)\widetilde G(\p-\q) \delta_{ab}\, ,
  \label{eq:Sigma}
\end{eqnarray}
where the dashed line corresponds 
to the sum of bubble chain diagrams, which is denoted by 
$\widetilde I$:
\begin{eqnarray}
  \widetilde I(\p) &=&
\;
\raisebox{-2mm}{\includegraphics[width=1.8cm]{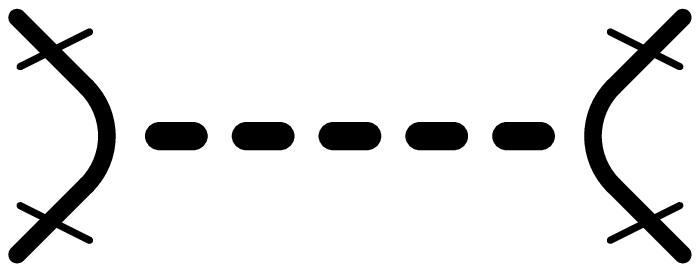}}
\;
\nonumber \\
   &=&
\;
\raisebox{-4mm}{\includegraphics[width=10mm]{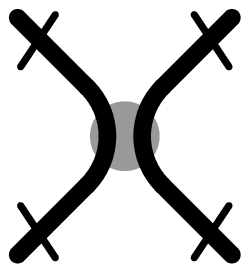}}
\;
 + \cdots +
\;
\raisebox{-3mm}{\includegraphics[width=2.5cm]{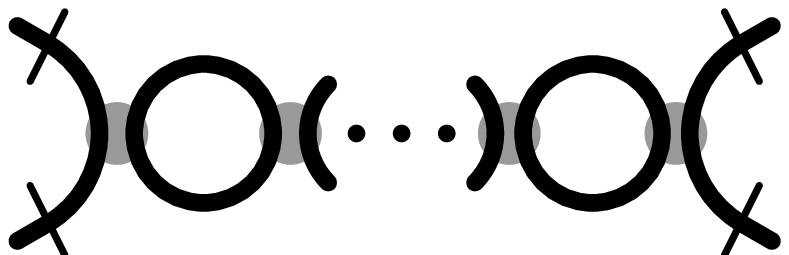}}
\;
\ \ \ +\ \ \cdots 
\nonumber \\
   &=& \frac{\lambda }{3N} 
       \frac{1}{1+\frac{\lambda }{6}\wPi(\p)}\, ,
  \label{eq:I}
\end{eqnarray}
with the one-loop polarization function
\begin{eqnarray} 
  \wPi (\p)&\equiv &\int \frac{d^3q}{(2\pi) ^3} 
  \widetilde G(\p-\q)\widetilde G(\q).
\end{eqnarray}
One should keep in mind that the vertex $\widetilde I$ is an 
${\cal O}(1/N)$ quantity. With the NLO self-energy (\ref{eq:Sigma}),
the KB equation for 
$\widetilde G_{ab}^{-1}(\p)=\widetilde G^{-1}(\p) \delta_{ab}$ reads 
 \begin{eqnarray}
 \widetilde G^{-1}(\p)
=\widetilde G_{0}^{-1}(\p)
+\frac{\lambda}{6}
\int \frac{d^3q}{(2\pi)^3} \widetilde G(\q)
+ 
\int \frac{d^3q}{(2\pi)^3} \widetilde I(\q)\widetilde G(\p-\q)
\; .
  \label{eq:sym KB; mom}
 \end{eqnarray}

\subsection{Calculation of $\eta$}

In Ref.~\cite{ABC} the critical exponent $\eta $ was calculated 
from the KB equation (\ref{eq:sym KB; mom}) at the critical point. 
Let us briefly review here how to obtain $\eta$, which is also necessary for 
the calculation of $\nu$.

Recall that $\widetilde G(\p)$ should become massless at the 
critical point: $\widetilde G^{-1}(\0)=0$. 
We can make this condition explicit for 
the KB equation (\ref{eq:sym KB; mom}) by subtracting 
the corresponding equation evaluated at $\p =0$. 
Then, the KB equation
at the critical point reads:
\begin{align}
\widetilde G^{-1}(\p)
= \p^2+ 
\int \frac{d^3q}{(2\pi)^3} \widetilde I(\q)
\left[\widetilde G(\p-\q)-\widetilde G(\q)\right] \, .
\label{eq:sym KB; crit}
\end{align}

\begin{figure}[t]
 \begin{center}
  \includegraphics[width=0.7\textwidth]{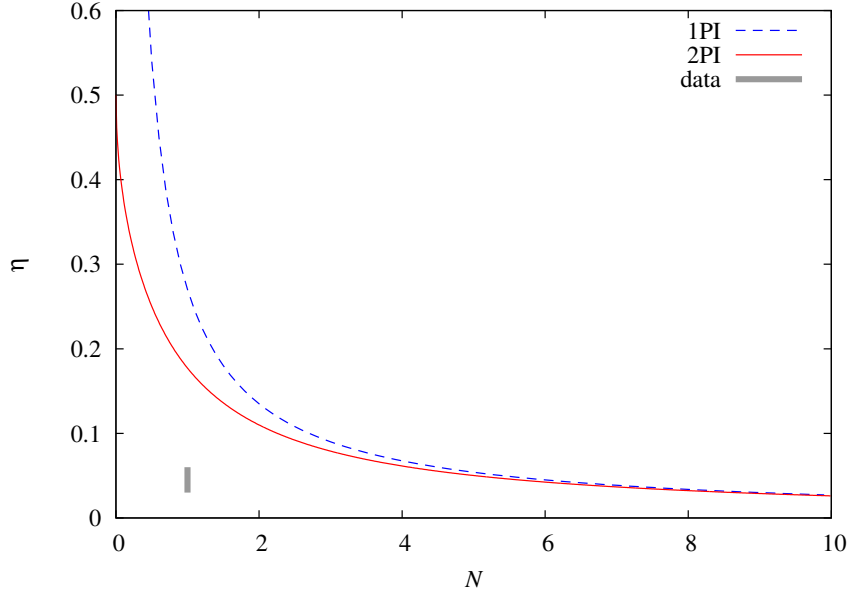}
 \end{center}
 \caption{The critical exponent $\eta$ as a function of $N$.
 The 2PI (1PI) result is shown in a solid (dashed) curve. }
 \label{eta graph}
\end{figure}

As one approaches the critical point, the polarization $\wPi$ dominates
in the denominator in the scaling region, 
and therefore we can ignore ``1''
in the denominator of $\wI(\p)$ on the right-hand side of Eq.~(\ref{eq:I}):
 \begin{eqnarray}
 \widetilde I (\p) 
  &\sim& \frac{2}{N} \wPi^{-1} (\p) \, .
 \label{eq:Iasympt}
 \end{eqnarray} 
Indeed, in order to investigate the asymptotic behavior of 
Eq.~(\ref{eq:sym KB; crit}) in the small momentum region, 
we can use the scaling form for $\widetilde G(\p)$ 
\begin{eqnarray}
  \widetilde G(\p)=
  \frac{1}{p^2} \left(\frac{p}{\Lambda }\right)^{\eta}\, 
  \label{eq:G_cp}
\end{eqnarray}
with $p=|\p|$ and a cutoff scale $\Lambda$, 
and find that $\wPi (\p)$ is infra-red singular
as long as $\eta < 1/2$:
\begin{eqnarray}
 \wPi (\p) ={\cal A} (\eta )\, \frac{1}{p}  
 \left (\frac{p}{\Lambda} \right )^{2 \eta }\, ,
  \label{eq:Pi_cp}
\end{eqnarray}
where
 \begin{eqnarray}
  {\cal A}(\eta )&=&
\frac{1}{8\pi ^{3/2}} 
\frac{\Gamma \left( \frac{1}{2} -\eta \right) 
      \left[ \Gamma \left( \frac{1+\eta }{2} \right)\right] ^2}
     {\left[ \Gamma \left( 1-\frac{\eta }{2} \right)\right] ^2 
             \Gamma \left( 1+\eta \right)} 
\; .
\end{eqnarray}
After performing the remaining integral with the use of the scaling 
form (\ref{eq:G_cp}) and rescaling with $\Lambda$ of the scaling region,
the KB equation (\ref{eq:sym KB; crit}) reduces to
\begin{eqnarray}
  p^{2-\eta } &=& p^2+{\cal B}(\eta )p^{2-\eta }+F_{\eta }(p^2)\, , 
\label{eq:KB-eta}
\end{eqnarray}
where
\begin{eqnarray}
  {\cal B}(\eta ) &=&
  \frac{4\eta (1-2\eta ) \cos (\eta \pi )}
       {(3-\eta )(2-\eta ) \sin ^2 (\eta \pi /2)N}
\nonumber
\end{eqnarray}
and
\begin{eqnarray}
  F_{\eta } &=&
 -\frac{(1-\eta )(2-\eta )}
       {6\pi ^2 \eta {\cal A}(\eta )N} p^2
 +O\left( p^4\right) \, .\nonumber
 \end{eqnarray}
In Eq.~(\ref{eq:KB-eta}), 
terms with $p^{2-\eta }$ are dominant at small momentum, $p\sim 0$, 
and determine the long distance behavior.
Equating the coefficients of $p^{2-\eta }$, one observes that $\eta$ has
to satisfy
\begin{eqnarray}
  1={\cal B}(\eta ) \; .
\label{eq:eta}
\end{eqnarray}
This gives the NLO result of $\eta$ in the 2PI $1/N$ expansion. 
In Fig.~\ref{eta graph} we show  the exponent $\eta$  
fixed by Eq.~(\ref{eq:eta}) as a function of $N$ in a solid line,
and compare it with the 1PI result written in a dashed line.
It is evident that the divergence of $\eta$ at $N=0$, 
which is seen in the 1PI result, is now resolved in the 2PI result,
and that the 2PI result is closer to the experimental values \cite{Chaikin}. 

\subsection{Calculation of $\nu$}

\subsubsection{Self-consistent equation for $\Gamma^{(2,1)}$ and iterative solution}

Now, we proceed to the calculation of $\nu$ in the 2PI 
formalism up to the NLO in the $1/N$ expansion. As we explained 
in Sec.~II, we will compute $\nu$ from the three-point vertex 
function $\widetilde \Gamma^{(2,1)}(\k/2,\k/2;\k)$,
which satisfies the self-consistent equation (\ref{eq:mom}).
Thus, the first thing to do is to determine
the NLO form of the kernel $D$ in Eq.~(\ref{eq:mom}). 
According to the definition of $D$ in Eq.~(\ref{eq:D4C}), 
it is obtained by differentiation of the self-energy $\Sigma$ 
with respect to $G$
in the NLO approximation and is written explicitly as
\begin{align}
\hspace{-1cm} \widetilde D(\p,\q;\p',\q') 
=& 
\;
\raisebox{-5mm}{\includegraphics[width=15mm]{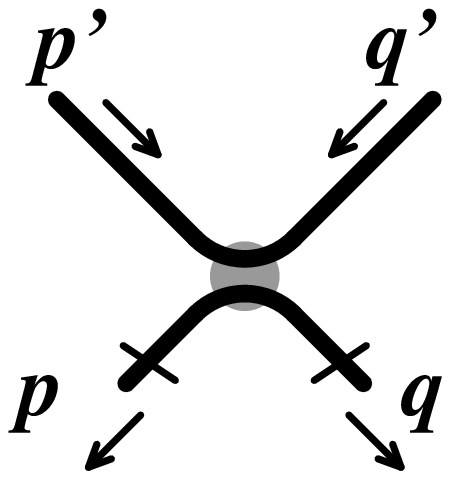}}
\;
  +
\;
\raisebox{-5mm}{\includegraphics[width=20mm]{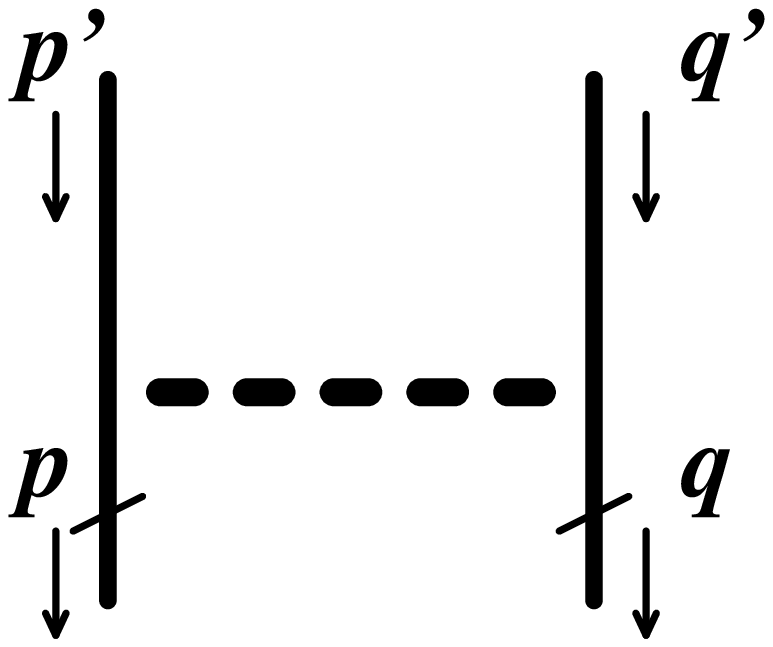}}
\;
  +
\;
\raisebox{-5mm}{\includegraphics[width=20mm]{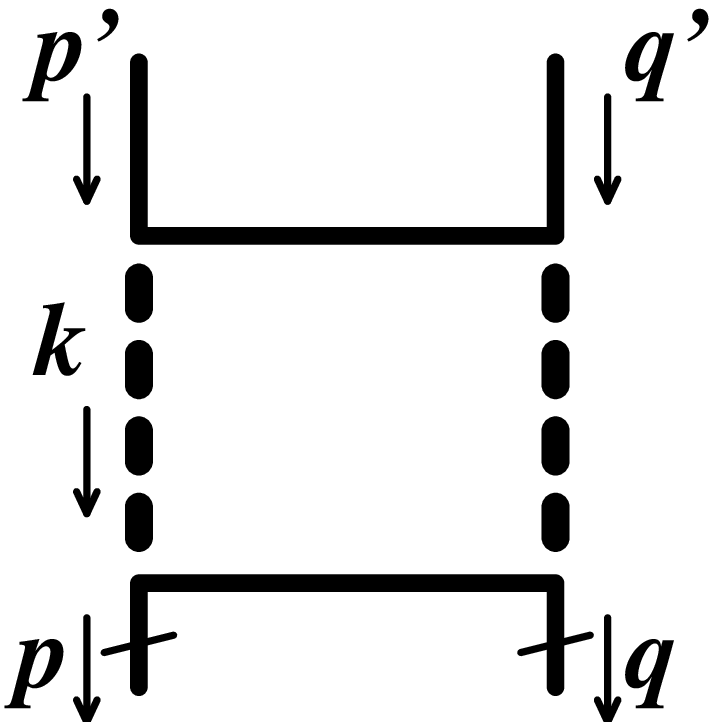}}
\;
\nonumber \\
=&
\left ( -\frac{\lambda }{6} -  \widetilde I (\p'-\p)
+N \int \frac{d^3 k}{(2\pi)^3} 
    \widetilde G(\k-\p) \widetilde G(\k-\p') 
    \widetilde I(\k) \widetilde I(\k-\p-\q)
\right )
\nonumber \\
& \times  \; \widetilde G(\p')\widetilde G(\q') 
           \; (2\pi)^3 \delta (\p+\q-\p'-\q')
\nonumber \\
\equiv & D_0+\frac{1}{N}D_1 
\, .
\label{eq:s.e.D}
\end{align}
The first term is of ${\cal O}(1)$, which is obtained from 
the first diagram in Eq.~(\ref{eq:Sigma}) by cutting the loop.
The second and third diagrams are of ${\cal O}(1/N)$ (recall that 
$\widetilde I$ is of ${\cal O}(1/N)$), which are 
obtained from the second diagram in Eq.~(\ref{eq:Sigma}) 
by cutting the solid  and dashed lines in the loop, respectively.
We introduced by the last equality the shorthand notations 
$D_0$ and $\frac{1}{N}D_1$ , respectively, 
for ${\cal O}(1)$ and ${\cal O}(1/N)$ contributions.

When this decomposition is substituted, the self-consistent equation 
(\ref{eq:mom}) for $\widetilde \Gamma^{(2,1)}$
is now expressed symbolically as
\begin{align}
  \Gamma^{(2,1)} &= 1 + \left(D_0 + \frac{1}{N}D_1\right)\Gamma^{(2,1)}, 
  \label{eq:Gamma(2,1):eq1}
\end{align}
which is diagrammatically represented as
 \begin{eqnarray}
\raisebox{-10mm}{\includegraphics[width=20mm]{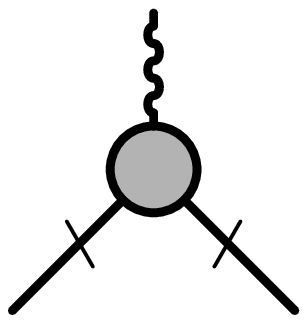}}
\;
  &=&
\;
\raisebox{-10mm}{\includegraphics[width=20mm]{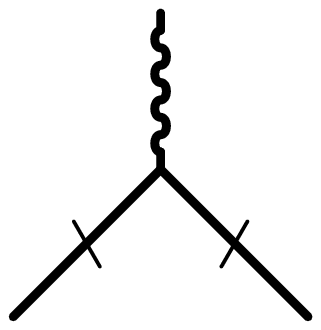}}
\;
+
\;
\raisebox{-10mm}{\includegraphics[width=20mm]{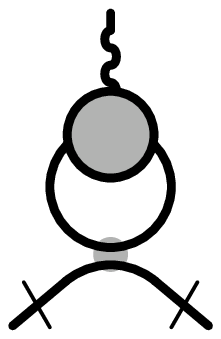}}
\;
+
\;
\raisebox{-10mm}{\includegraphics[width=20mm]{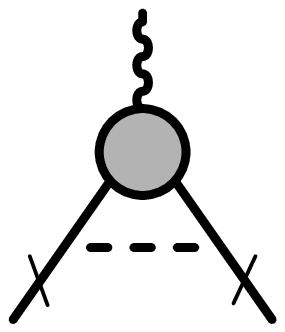}}
\;
+
\;
\raisebox{-10mm}{\includegraphics[width=20mm]{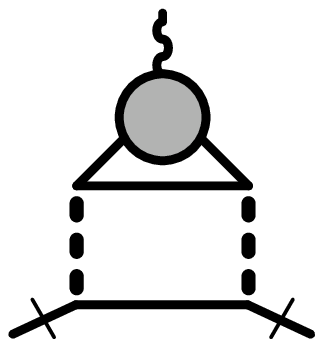}}
\;
\, .
   \label{eq:Gamma(2,1):graph}
 \end{eqnarray}
Here, the first and second terms are of ${\cal O}(1)$, while the 
third and fourth terms are of ${\cal O}(1/N)$.

In the present paper, we will not solve 
Eq.~(\ref{eq:Gamma(2,1):eq1}) self-consistently, but
we will follow the procedure of Ref.~\cite{Ma}
in the evaluation of the critical exponent $\nu$. 
That is, we extract contributions logarithmically divergent at 
small momentum, $\ln k$, from the LO and NLO terms in 
$\widetilde \Gamma^{(2,1)}(\k/2,\k/2 ; \k)$, which will give the 
critical exponent $\nu$ when exponentiated.

Notice that at the LO the equation, 
$\Gamma^{(2,1)} = 1  +D_0 \Gamma^{(2,1)}$, is immediately 
solved by $\Gamma^{(2,1)}_{\rm LO}=1/ (1-D_0 )$
with $D_0 (\p)= - (\lambda/6)\wPi(\p)$, 
which is nothing but the sum of 
bubble chain diagrams proportional to 
$\widetilde I(\p)$ in Eq.~(\ref{eq:I}).
This is easily understood from the ${\cal O}(1)$ diagrams 
shown in Eq.~(\ref{eq:Gamma(2,1):graph}).
We will denote $1/(1-D_0)$ with the same dashed line as $\widetilde I$
in the diagrams. From the low momentum behavior of 
$\widetilde \Gamma^{(2,1)}_{\rm LO}(\k/2, \k/2; \k)=1/ (1-D_0(\k) )$,
one should be able to get the critical exponent $\nu$ at the LO.
Indeed, it gives rise to
 \begin{align}
\widetilde \Gamma^{(2,1)}_{\rm LO}(\k/2, \k/2; \k)
&=1/ (1-D_0(\k) )  \nonumber \\
&\sim \frac{6}{\lambda } \widetilde \Pi ^{-1} (\k)
 =\frac{6}{\lambda {\cal A}(\eta )}
\;  k \left ( \frac{k}{\Lambda} \right )^{-2\eta}
\;  ,
\label{eq:(a)}
\end{align}
where the polarization $\Pi$ is evaluated with the scaling form for 
$\widetilde G(\p)$. 
There is no $\ln k$ dependence in this result, but rather
it directly gives the exponent $\nu$  at the LO. 
Comparing this with the scaling behavior 
$\widetilde \Gamma^{(2,1)} \sim k^{2-\eta-1/\nu}$ and using the LO result for 
$\eta$, {\it i.e.}, $\eta_{\rm LO}=0$,  we find that $\nu$ at the LO is
\begin{align}
\nu^{}_{\rm LO} =1\, .\label{nu-LO}
\end{align}
This is the well-known result in the $1/N$ expansion analysis.
Non-trivial correction for $\nu$ should be obtained at the NLO.

With $\Gamma^{(2,1)}_{\rm LO}=1/(1-D_0)$,
we can rewrite Eq.~(\ref{eq:Gamma(2,1):eq1}) in a form which is 
more suitable for the perturbation expansion:
\begin{align}
\Gamma^{(2,1)}
&=
\frac{1}{1-D_0}  + \frac{1}{1-D_0}\frac{1}{N}D_1\Gamma^{(2,1)}
\nonumber\\
&=
\frac{1}{1-D_0} + 
\left(1 + \frac{1}{1-D_0}D_0\right)\frac{1}{N}D_1\Gamma^{(2,1)}
\nonumber \\
& =
\frac{1}{1-D_0} + 
\left(1 + \frac{1}{1-D_0}D_0\right)\frac{1}{N}D_1
\frac{1}{1-D_0} +\cdots
\; ,
\label{eq:Gamma(2,1):eq2}
\end{align}
where on the third line we have solved the equation 
iteratively and shown only the LO and NLO contributions explicitly.
Diagramatically this iterative solution is represented as
\begin{eqnarray}
\;
\raisebox{-10mm}{\includegraphics[width=18mm]{Gamma.eps}}
\;
  &=&
\;
\raisebox{-10mm}{\includegraphics[width=18mm]{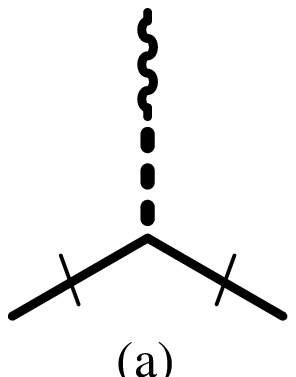}}
\;
   +
\;
\raisebox{-10mm}{\includegraphics[width=18mm]{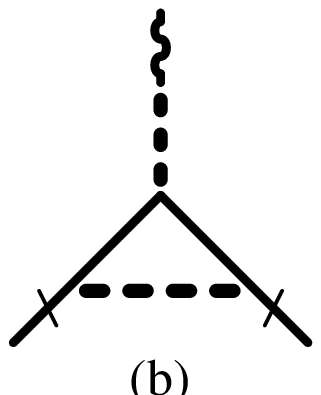}}
\;
   +
\;
\raisebox{-10mm}{\includegraphics[width=18mm]{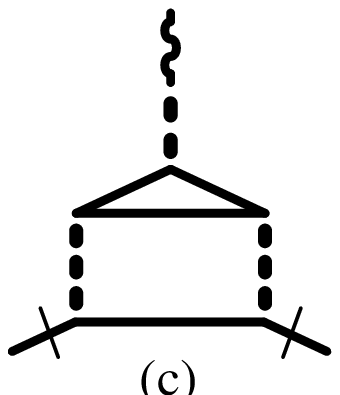}}
\;
   +
\;
\raisebox{-10mm}{\includegraphics[width=18mm]{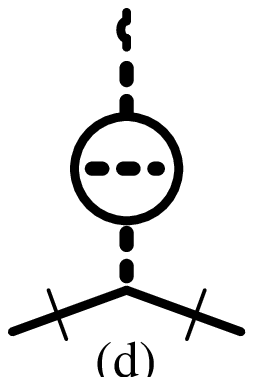}}
\;
   +
\;
\raisebox{-10mm}{\includegraphics[width=18mm]{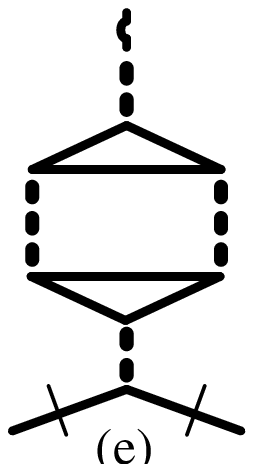}}
\;
 + \cdots
  \label{eq:iteration}
 \end{eqnarray}
The diagrams (b) and (c) correspond to the contributions
from $1$,
 and the last two (d) and (e) correspond to the contributions from 
$\frac{1}{1-D_0} D_0$ in the parenthesis of Eq.~(\ref{eq:Gamma(2,1):eq2}). 
We note that $\tfrac12\varphi^2$ operator is first
attached to $\Gamma^{(2,1)}_{\rm LO}=1/(1-D_0)$ in all the diagrams.

\subsubsection{Evaluation of each diagram}

We now evaluate the ${\cal O}(1/N)$ contributions to 
$\Gamma^{(2,1)}$ as shown in Eq.~(\ref{eq:iteration}) 
by using the scaling form (\ref{eq:G_cp}) for 
$\widetilde G(\p)$. One should note here that the use of 
the scaling form is a non-perturbative prescription.
We will substitute in $\widetilde G(\p)$ 
the exponent $\eta$ obtained by
the self-consistent KB equation at the NLO .

Let us examine the four NLO diagrams (b), (c), (d) and
 (e) in Eq.~(\ref{eq:iteration}) one by one, 
seeking for the $\ln k$ dependence.
We note that the common factor $\Gamma^{(2,1)}_{\rm LO}$ attached to $\tfrac12\varphi^2$ 
operator in all the four diagrams gives 
$1/(1-D_0) \sim ( 6/ \lambda {\cal A}(\eta )) k^{1-2\eta }$.
The remaining part in each diagram will result in the $\ln k$ dependence 
to modify the exponent $\nu$.
In this subsection we deal with only diagrams (b) and (d) because  
the contributions from other diagrams (c) and (e) are negligibly small as 
explained in Appendix.

\begin{figure}[t]
\includegraphics[width=30mm]{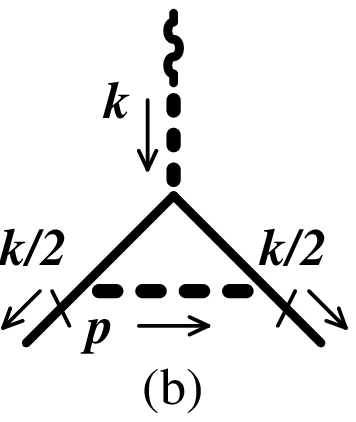}
\hfil
\includegraphics[width=30mm]{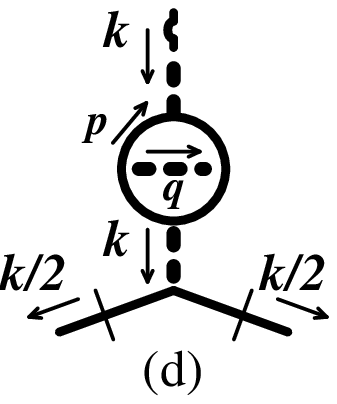}
 \caption{NLO diagrams (b) and (d) contributing to the exponent $\nu$,
with explicit momentum assignment.}
 \label{NLO-diagrams}
\end{figure}

Firstly, together with the asymptotic form for $\widetilde I$ 
(\ref{eq:Iasympt}), diagram (b) is evaluated as 
 \begin{align}
 {\rm (b)}\ 
 \sim&
 \frac{6}{\lambda {\cal A}(\eta )} k^{1-2\eta }
       \int \frac{d^3p}{(2\pi) ^3}\, 
     \widetilde G ( \p+ {\k}/{2} )\, \widetilde G( \p-{\k}/{2})\, 
       \left(\frac{-2}{N}\widetilde \Pi ^{-1} (\p)\right)  \nonumber \\
 \sim&
 \frac{6}{\lambda {\cal A}(\eta )} k^{1-2\eta } 
\left[\frac{-2}{N} \int^\Lambda \frac{d^3p}{(2\pi) ^3} 
      \left\vert \p+\k/2 \right\vert ^{-2+\eta } 
      \left\vert \p-\k/2 \right\vert ^{-2+\eta } 
              p ^{1-2\eta } {\cal A}(\eta )^{-1} \right]\, .
  \label{eq:(b)}
 \end{align}
See Fig.~\ref{NLO-diagrams}, for the assignment of each 
momentum. 
The minus sign in 
$\frac{-2}{N}\widetilde \Pi^{-1}(\p)\sim -\widetilde I(\p)$ 
originates from that of the second term in Eq.~(\ref{eq:s.e.D}). 
We have introduced an upper cutoff $\Lambda$ of the scaling momentum region,
while we have omitted $\Lambda^\eta$ factors, which can be easily restored.
Notice that this integral is logarithmically divergent when $k=0$, which 
indicates the infra-red dominance in the $p$-integration and justifies
the use of the scaling form for $\widetilde G(\p)$.
Indeed, the integral yields the $\ln k$ contribution as 
\begin{eqnarray}
{\rm (b)} 
 &\sim & \frac{6}{\lambda {{\cal A}(\eta)}} k^{1-2\eta } 
         \left[\frac{-2}{N{\cal A}(\eta)} \int^\Lambda_k 
         \frac{d^3p}{(2\pi) ^3} 
           p ^{-2+\eta } p ^{-2+\eta } p ^{1-2\eta }
         \right] \nonumber \\
 &\sim & \frac{6}{\lambda {{\cal A}(\eta)}} k^{1-2\eta } 
         \left[ \frac{1}{\pi ^2 N{{\cal A}(\eta)}} \ln k \right] ,
  \label{eq:(b)result}
 \end{eqnarray}
where we have picked up only the most singular part in $k \to 0$.

Secondly, diagram (d) 
(see Fig.~\ref{NLO-diagrams} for momentum assignment)
is evaluated in a similar way as
\begin{eqnarray}
 {\rm (d)}
&\sim & \left ( \frac{6}{\lambda {\cal A}(\eta )} k^{1-2\eta } \right )^2
    \left (\frac{-\lambda}{6} \right )
      \int \frac{d^3p \, d^3q}{(2\pi) ^6}
      \widetilde G(\p) \widetilde G( \p+\k) 
      \left(\frac{-2}{N}\widetilde \Pi ^{-1} (\q)\right) 
      \widetilde G( \p+\q) \widetilde G(\p+\q+\k) 
\nonumber \\
&\sim & \frac{6}{\lambda {\cal A}(\eta )} k^{1-2\eta }
    \left[ \frac{k^{1-2\eta }}{{\cal A}^2(\eta)} 
      \frac{2}{N}
      \int^\Lambda  \frac{d^3p \, d^3q}{(2\pi) ^6} \;
       p^{-2+\eta } \vert \p+\k \vert^{-2+\eta } 
       \vert \p+\q\vert^{-2+\eta } \vert \p+\q+\k\vert^{-2+\eta }
       q^{1-2\eta }
    \right] \nonumber \\
&=& \frac{6}{\lambda {\cal A}(\eta )}k^{1-2\eta }
    \left[  \frac{2}{N{\cal A}(\eta)^2}
      \int^{\Lambda /k} \frac{d^3u\, d^3v}{(2\pi) ^6}\; 
       u ^{-2+\eta } \vert \u+\hat \k\vert^{-2+\eta } 
      \vert \u + \v\vert^{-2+\eta } \vert\u + \v + \hat \k\vert^{-2+\eta }  
      v ^{1-2\eta }
    \right]\, ,
\nonumber
\end{eqnarray}
where we have rescaled the variables 
$\u = \p / |\k|$, $\v= \q / |\k|$ and $\hat \k=\k/|\k|$. 
The $\ln k$ dependence comes from two regions of the above integral:
(I)  $|\v| \sim \Lambda/k$, $|\u| \ll |\v|$ and (II) $|\u| \sim |\v| \sim \Lambda/k$, $|\u+\v| \ll |\u|,\ |\v|$.
One notices that the change of the variables, $\u' = \u+\v$ and $\v' = -\v$, maps region II to region I and vice versa with keeping the integral the same, which means that these two regions give the same $\ln k$ contributions.
Therefore, the $\ln k$ contribution in the above integral coincides with twice that from region I:
 \begin{eqnarray}
  {\rm (d)}
&\sim& \frac{6}{\lambda {\cal A}(\eta )}k^{1-2\eta }
    \left[  \frac{2}{N{\cal A}(\eta)^2}
      2\times \int_{\rm I} \frac{d^3u\, d^3v}{(2\pi) ^6}\; 
       u ^{-2+\eta } \vert \u+\hat \k\vert^{-2+\eta } 
      \vert \u + \v\vert^{-2+\eta } \vert\u + \v + \hat \k\vert^{-2+\eta }  
      v ^{1-2\eta }
          \right] \nonumber  \\
 &\sim &  \frac{6}{\lambda {\cal A}(\eta )}  k^{1-2\eta } 
        \left[ 
          \frac{4}{N{{\cal A}(\eta)}^2}
      \int^{\Lambda/k} \frac{d^3u\, d^3v}{(2\pi) ^6}
 \,     u^{-2+\eta } \vert \u+ \hat \k \vert^{-2+\eta } v ^{-3}
        \right] \nonumber \\
      &=& \frac{6}{\lambda {\cal A}(\eta )}  k^{1-2\eta } 
        \left[
          \frac{4}{N{{\cal A}(\eta)}^2} {{\cal A}(\eta)}
          \int^{\Lambda/k} \frac{d^3v}{(2\pi) ^3} \,  v ^{-3}
        \right] \nonumber \\
      &\sim& \frac{6}{\lambda {\cal A}(\eta )}  k^{1-2\eta } 
         \left[
          \frac{-2}{\pi ^2 N{{\cal A}(\eta)}} \ln k 
         \right] .
  \label{eq:(d)result}
 \end{eqnarray}

Finally, we checked that both the coefficients in 
the diagrams (c) and (e) in Eq.~(\ref{eq:iteration}) are 
consistent with zero in a numerical integration.
Therefore, we do not include these two diagrams as alluded before. 
Details of the calculation are shown in Appendix.
Notice that these two diagrams do not contribute to $\nu$ 
in the 1PI $1/N$ expansion \cite{Ma}.

\subsubsection{Result}

\begin{figure}[t]
 \begin{center}
  \includegraphics[width=0.7\textwidth]{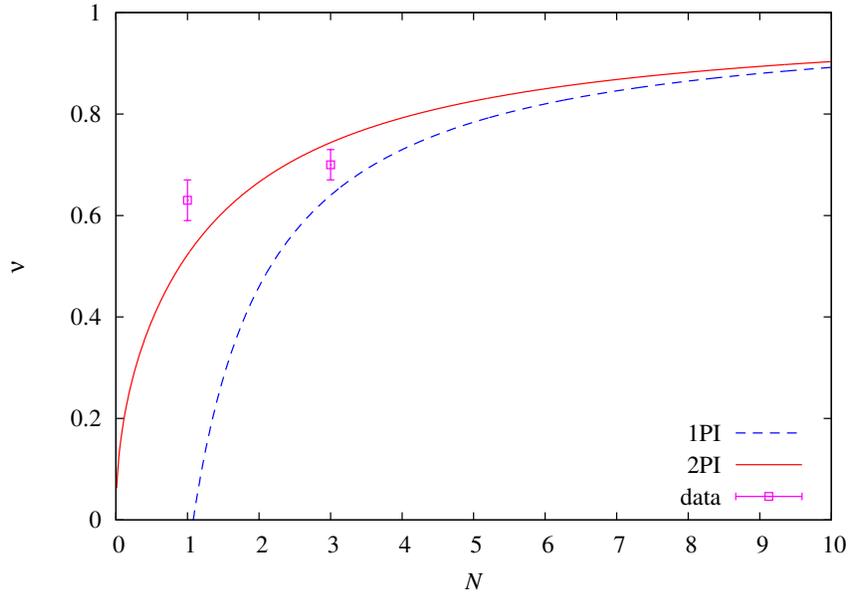}
 \end{center}
\caption{
Exponent $\nu$ from 2PI NLO calculation as a function of $N$ (solid line). 
The result from 1PI NLO calculation (dashed line) and experimental data \cite{Chaikin} 
are shown for comparison. 
}
 \label{nu2 graph}
\end{figure}

We collect the NLO corrections  (\ref{eq:(b)result}) 
and (\ref{eq:(d)result}) to the LO result (\ref{eq:(a)}).
As we noticed before, these NLO contributions (\ref{eq:(b)result}) 
and (\ref{eq:(d)result}) have the same prefactor $\widetilde \Gamma^{(2,1)}_{\rm LO}$.
Therefore, we are able to exponentiate the $\ln k$ term 
to obtain $\widetilde \Gamma ^{(2,1)}$ at the NLO: 
\begin{eqnarray}
       \widetilde \Gamma^{(2,1)}_{\rm NLO} \left(\frac{\k}{2},\frac{\k}{2};\k \right)
&\sim& \widetilde \Gamma^{(2,1)}_{\rm LO}\left(\frac{\k}{2},\frac{\k}{2};\k \right)
    \left[ 1-\frac{1}{\pi ^2 N{\cal A}(\eta)}\ln k \right]  \nonumber \\
&\sim & \frac{6}{\lambda {\cal A}(\eta )} \ k^{1-2\eta-\frac{1}{\pi ^2N{\cal A}(\eta)}}\, .
 \end{eqnarray}
Then, comparing this with Eq.~(\ref{eq:Gamma^(2,1)}), one finally obtains
the exponent $\nu$ of the 2PI NLO calculation:
 \begin{eqnarray}
  \nu^{\rm (2PI)}_{\rm NLO} 
   = \frac{1}{1+\eta + \frac{1}{\pi ^2N{\cal A}(\eta)}}\, ,
  \label{1/N result}
 \end{eqnarray}
where $\eta$ should be the 2PI critical exponent (\ref{eq:eta})
for consistency. 
This is our main result. 
In the limit $N\to \infty$, this result of course recovers the LO result
(\ref{nu-LO}) together with $\lim_{N\to \infty} \eta(N)= 0$.
We plot $\nu$ as a function of $N$ in Fig.~\ref{nu2 graph},
where $\nu$ from the standard 1PI action at the NLO \cite{Ma}
 \begin{eqnarray}
  \nu^{\rm (1PI)}_{\rm NLO}  =1-\frac{3\pi^2}{32 N}
\label{1PI-NLO}
 \end{eqnarray}
 is also shown for comparison. 
At large $N$ the difference between the 1PI and 2PI results diminishes
and both converge to the LO result $\nu_{\rm LO}=1$. At small $N$
the 2PI result stays positive, while the 1PI result can become negative. 
We see that the result of the 2PI NLO calculation gives an improved estimate 
for the exponent $\nu$ at $N=1, 3$ than the 1PI NLO result.

\section{summary and discussion}
In the present paper we have developed a method to compute 
the critical exponent $\nu$ using the 2PI effective action. 
Although $\nu$ is associated with the diverging behavior of the 
correlation length $\xi$ near the critical point, 
one can compute it on the critical point by analyzing 
the three-point vertex function $\Gamma ^{(2,1)}$. 
Roughly speaking, this is possible because 
$\Gamma^{(2,1)}$ is a derivative of the correlation 
function with respect to the temperature, and thus includes 
certain information on the deviation from the critical point. 
In the 2PI formalism we can write down a self-consistent 
equation for $\Gamma^{(2,1)}$, which is easily derived from 
the KB equation for the two-point function. The explicit form of the 
equation was obtained to the NLO in the $1/N$ expansion. 
We solved this equation by iteration to the NLO in the $1/N$ expansion,
and identified from the resultant $\Gamma^{(2,1)}$ the exponent $\nu$,
Eq.~(\ref{1/N result}), as shown in Fig.~\ref{nu2 graph}.

The difference between the 2PI NLO result (\ref{1/N result}) and 
the 1PI NLO result (\ref{1PI-NLO}) comes from two points as follows:
Firstly, in the 2PI formalism, we deal with the full propagator 
in contrast to the free propagator in the 1PI formalism. 
Secondly, 
the sets of the NLO diagrams for $\Gamma^{(2,1)}$ 
are different between the 1PI and 2PI formalisms, 
while there is only one LO diagram which is common in both.
Namely, the 1PI NLO calculation involves the following five diagrams \cite{Ma} 
which should be compared with four 2PI diagrams shown in 
Eq.~(\ref{eq:iteration}): 
 \begin{eqnarray}
\includegraphics[width=20mm]{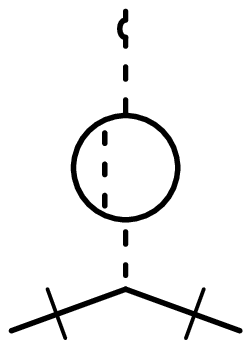}
\quad
\includegraphics[width=20mm]{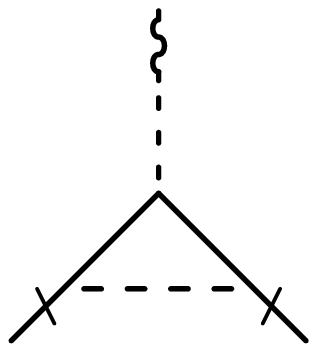}
\quad
\includegraphics[width=20mm]{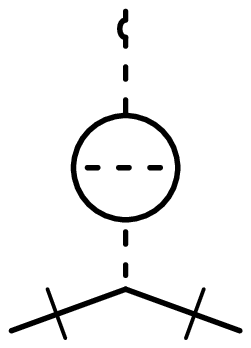}
\quad
\includegraphics[width=20mm]{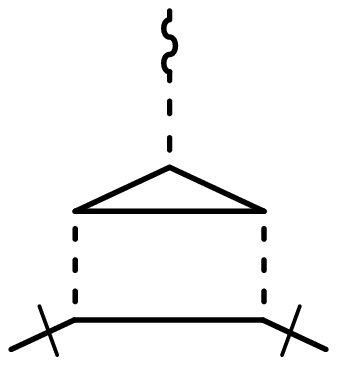}
\quad
\includegraphics[width=20mm]{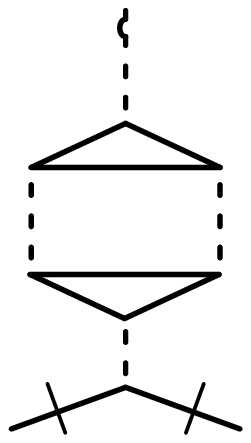}
  \label{eq:NLO 1PI}
 \end{eqnarray}
There is no self-energy insertion in the 2PI diagrams because it is
already resummed in the full propagator.
In fact, the first 1PI diagram contains one self-energy insertion,
and is already included in the 2PI LO diagram
Eq.~(\ref{eq:iteration}) (a).
This resummation of the self-energy diagrams in the 2PI formalism enables us
to take account of important higher-order contributions into the form of the full propagator,
which is the origin of the improvement of the 2PI result over the 1PI result in the
calculation of the critical exponent.

Notice that the 1PI NLO result (\ref{1PI-NLO}) has an apparent 
flaw at small $N$. 
The exponent $\nu$ must be positive $\nu>0$ as it describes 
the diverging behavior of the correlation length $\xi$
near the critical point ({\it cf.} Eq.~(\ref{eq:xi})). 
However, the 1PI NLO result becomes
negative at small $N$, although such a small value of $N$ is 
outside the validity region of the $1/N$ expansion in a strict sense.
In contrast, the exponent $\nu$ in the 2PI NLO result
remains positive for all $N$,
and it is closer to the experimental data
at $N=1,3$ \cite{Chaikin}.

Expanding $\eta $ of Eq.~(\ref{1/N result}) in $N$, we see that the
2PI result reproduces  Ma's 1PI result (\ref{1PI-NLO}) \cite{Ma}, 
and includes a part of higher order terms. 
It shows that the 2PI effective action resums not only leading-log-terms which
are resummed in the 1PI calculation but also a certain class of the higer-log-terms.

In the present paper, we did not require the self-consistency for
$\Gamma^{(2,1)}$, but rather solved the equation by iteration to the NLO.
Using conformal invariance in the coordinate space
\cite{Vasil'ev,Vasil'ev et,Polyakov} in (\ref{eq:coord}) at the critical point, 
one may analyze the self-consitent solution for $\Gamma^{(2,1)}$
to get a better estimate for $\nu$. However, one should keep in mind that
the higher-order calculation of the exponent $\nu$ in the 
standard 1PI formalism
up to ${\cal O}(1/N^2)$ \cite{Justin,AbeHikami} tends to deviate from the experimental
values. 
Inclusion of higher-order terms by requiring the self-consistency 
for $\Gamma^{(2,1)}$
is an open issue.

\section*{Acknowledgements}
This work was initiated at the workshop 
``{\it Non-equilibrium quantum field theories and dynamic critical phenomena}''
(YITP-T-08-07) at YITP, Kyoto Univeristy, 2009.
The authors are grateful to J\"urgen Berges who drew their 
attention to Ref.~\cite{ABC} and stimulating discussions during the workshop.
They also thank Hiroyuki Kawamura for discussions on renormalization issues.
One of the authors (H.F.) acknowldges warm hospitality of Technische Universit\"at Darmstadt, where part of this work was done.

\section*{Appendix}

\begin{figure}
\includegraphics[width=30mm]{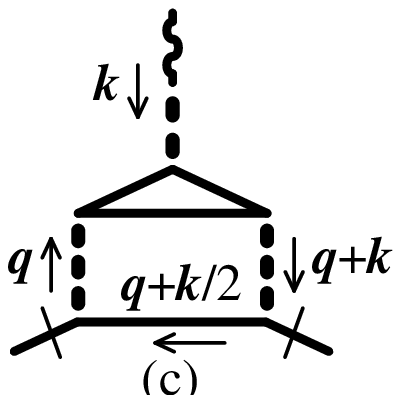}
\hfil
\includegraphics[width=30mm]{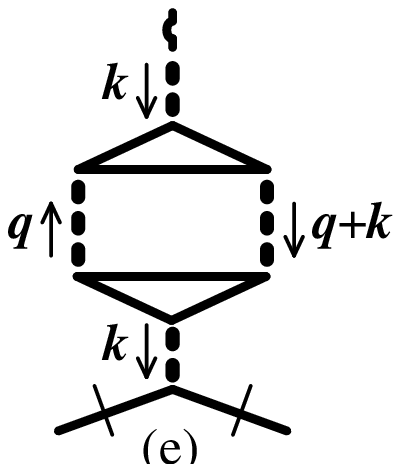}
\caption{NLO diagrams (c) and (e). }
\label{NLO-diagrams-ce}
\end{figure}

Here we show details of the calculation of diagrams (c) and (e) 
in Eq.~(\ref{eq:iteration}). We closely follow Ref.~\cite{Ma}
for identification and evaluation of the $\ln k$ contributions.
With the momentum assingment shown in Fig.~\ref{NLO-diagrams-ce}, 
each diagram is calculated as follows:
 \begin{eqnarray}
{\rm (c)}
  &\sim & \frac{6}{\lambda {\cal A}(\eta )} k^{1-2\eta } 
  \left[\frac{4}{N} \int \frac{d^3q}{(2\pi) ^3} 
      \widetilde \Pi ^{-1} (\q) \widetilde \Pi ^{-1} (\q+\k) 
      \widetilde G ( \q+ \k/2 ) 
      T(\k,\q)
  \right] ,
  \label{eq:(c)} \\
{\rm (e)}
&\sim& \frac{6}{\lambda {\cal A}(\eta )} k^{1-2\eta } 
   \left[ 
      \frac{4}{N} \int \frac{d^3q}{(2\pi) ^3} 
      \widetilde \Pi ^{-1} (\q) \widetilde \Pi ^{-1} (\q+\k) 
      T^2 (\k,\q) 
      \left(  \frac{-\lambda }{6} \cdot \frac{6}{\lambda } \Pi ^{-1} (\k) \right)
   \right] \nonumber \\
  &=& \frac{6}{\lambda {\cal A}(\eta )} k^{1-2\eta } 
      \left[ \frac{-4}{N} \frac{k^{1-2\eta }}{{\cal A}(\eta)}
      \int \frac{d^3q}{(2\pi) ^3} 
      \widetilde \Pi ^{-1} (\q)\widetilde  \Pi ^{-1} (\q+\k) 
     T^2 (\k,\q)
      \right] ,
  \label{eq:(e)}
 \end{eqnarray}
where $T(\k,\q)$ represents the triangle part and is defined by
 \begin{eqnarray}
  T(\k,\q) &\equiv& 
\raisebox{-10mm}{\includegraphics[width=25mm]{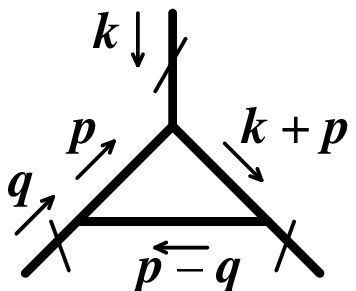}}
\nonumber \\
   &=& \int \frac{d^3p}{(2\pi) ^3} \,
       \widetilde G(\p)\, \widetilde G(\p+\k)\, 
       \widetilde G(\p-\q) \nonumber \\
   &\sim & \int \frac{d^3u}{(2\pi) ^3} \, 
       k^{-3+3\eta } \, {u} ^{-2+\eta } \, 
       \vert \u+\hat{\k}\vert^{-2+\eta } \, 
       \vert \u-\v\vert^{-2+\eta } \nonumber \\
   &\equiv& k^{-3+3\eta } \, T_0(\hat{\k},\v) .
  \label{def:T}
 \end{eqnarray}
Here we have again introduced dimensionless variables $\u=\p/|\k|$, 
$\v=\q/|\k|$, and $\hat\k =\k/|\k|$.
Then two diagrams become
 \begin{eqnarray}
  {\rm (c)} 
   &\sim & \frac{6}{\lambda {\cal A}(\eta )} k^{1-2\eta } 
        \left[\frac{4}{N} \int \frac{d^3v}{(2\pi) ^3} 
        v ^{1-2\eta } \vert\v+\hat{\k}\vert^{1-2\eta } 
        \vert \v +\hat{\k}/2 \vert ^{-2+\eta } \, 
         T_0(\hat{\k},\v) 
        \right],
  \label{eq:(c)result} \\
  {\rm (e)} 
      &\sim& \frac{6}{\lambda {\cal A}(\eta )}  k^{1-2\eta } 
        \left[ \frac{-4}{N{\cal A}(\eta)} \int \frac{d^3v}{(2\pi) ^3} 
      v ^{1-2\eta } \vert\v+\hat{\k}\vert^{1-2\eta } T_0^2 (\hat{\k},\v)
        \right]\, .
 \label{eq:(e)result}
 \end{eqnarray}
As we disscuss in the text, we are interested in the $\ln k$ 
contribution when $k$ is small. We first note that the 
$\ln k$ contribution appears not from the integration over $\u$ in $T_0$, 
but from the integration over $\v$ in $T_0$. 
Instead, it will appear from 
the integration over $\v$.  However, it is not straightforward 
to see the power of $\v$ from the above two expressions. 
If $T_0(\hat{\k},\v)$ generates $v ^{-3+3\eta }$, then
$\ln k$ terms appear in Eq.~(\ref{eq:(c)result});
 \begin{eqnarray}
  T_0(\hat \k,\v) \sim v ^{-3+3\eta } &\rightarrow & 
  {\rm (c)} \sim \int^{\Lambda/k} dv v ^{-1} \sim \ln k .
  \label{residue1} 
 \end{eqnarray}
Similarly, in Eq.~(\ref{eq:(e)result}), if 
$T_0^2(\hat{\k},\v)$ yields $v ^{-5+4\eta }$, 
then $\ln k$ terms appear;
 \begin{eqnarray}
  T_0^2(\hat{\k},\v) \sim v ^{-5+4\eta } &\rightarrow & 
{\rm (e)} \sim \int^{\Lambda/k} dv v ^{-1} \sim \ln k .
  \label{residue3} 
 \end{eqnarray}
Below we examine whether these powers indeed appear in $T$.
The definition of $T_0$ in Eq.~(\ref{def:T}) can be 
explicitly written as ($\theta$ and $\phi$ are the angles between
two vectors $(\u,\hat\k)$ and $(\u, \v)$, respectively)
 \begin{eqnarray}
  T_0(\hat{\k},\v) &=& \frac{1}{8\pi ^3} \int u ^2du \int d\Omega \, 
    u ^{-2+\eta } (u ^2+2u \cos \theta +1)^{-1+\eta /2} (u ^2+2u v \cos \phi +v ^2)^{-1+\eta /2} .
  \label{eq:polarT}
 \end{eqnarray} 
To see the power of $v$ in this quantity, it is convenient to 
perform the Mellin transformation defined by
 \begin{eqnarray}
  g(s)&=&\int _0 ^\infty dxf(x) x^{s-1},\\
  f(x)&=&\frac{1}{2\pi i} \int _{c-i\infty } ^{c+i\infty } ds x^{-s} g(s)\, .
 \end{eqnarray}
The $v$-dependent part of Eq.~(\ref{eq:polarT}) can be 
Mellin-transformed as follows:
 \begin{eqnarray}
  \int _0 ^\infty dv (u ^2+2u v \cos \phi +v ^2)^{-1+\eta /2} v ^{s-1}
  &=& 2^{1/2-\eta /2} (\sin \phi )^{\eta /2-1/2} \frac{\Gamma (s) \Gamma (2-\eta -s)}{\Gamma (2-\eta )} 
       \left( \frac{1+\cos \phi }{1-\cos \phi } \right) ^{\eta /4-1/4} \nonumber \\
   && \times {}_2 F_1 \left( -s-\frac{\eta }{2}+\frac{3}{2} ,s+\frac{\eta }{2}-\frac{1}{2},-\frac{\eta }{2}+\frac{3}{2},\frac{1-\cos \phi }{2} \right) \nonumber \\
   && \times u ^{s+\eta -2} \nonumber \\
   &\equiv & F(s,\eta ,\phi ) u ^{s+\eta -2}, \ \ \ \ \ (0<{\rm Re}[s]<2-\eta)
  \label{eq:after1Mellin}
 \end{eqnarray}
where $B$ is the beta function, $P_\mu ^\lambda $ is the 
associated Legendre function and $F$ is the hypergeometric function.
Here we have used the formula \cite{formula} ($0<{\rm Re}[s]<2\mu , \ -\pi <\theta <\pi $): 
 \begin{eqnarray}
  \int _0 ^\infty (x^2+2ax \cos \theta +a^2) ^{-\mu } x^{s-1} dx
   &=& 2^{\mu -1/2} (\sin \theta )^{1/2-\mu } \Gamma \left( \mu +\frac{1}{2} \right) B( s,2\mu -s) P_{s-\mu -1/2} ^{1/2-\mu } (\cos \theta ) a^{s-2\mu } \nonumber \\
   &=& 2^{\mu -1/2} (\sin \theta )^{1/2-\mu } \frac{\Gamma (s) \Gamma (2\mu -s)}{\Gamma (2\mu )} 
       \left( \frac{1+\cos \theta }{1-\cos \theta } \right) ^{1/4-\mu /2} \nonumber \\
   && \times {}_2 F_1 \left( -s+\mu +\frac{1}{2} ,s-\mu +\frac{1}{2},\mu +\frac{1}{2},\frac{1-\cos \theta }{2} \right) a^{s-2\mu }\, .
   \label{form}
 \end{eqnarray}
Therefore, Mellin transform of Eq.~(\ref{eq:polarT}) is given as 
 \begin{eqnarray}
  T_0(\hat{\k},s) &=& 
    \int _0 ^\infty dv T_0(\hat{\k},\v) v ^{s-1} \nonumber \\
    &=& \frac{1}{8\pi ^3} \int _0 ^\infty u ^2du \int d\Omega \, 
    u ^{-2+\eta } (u ^2+2u \cos \theta +1)^{-1+\eta /2} F(s,\eta ,\phi ) u ^{s+\eta -2} \nonumber \\
    &=& \frac{1}{8\pi ^3} \int d\Omega F(s,\eta ,\phi ) \int _0 ^\infty du u ^{s+2\eta -2} (u ^2+2u \cos \theta +1)^{-1+\eta /2} .
 \end{eqnarray}
If we use the formula (\ref{form}) again, then 
Eq.~(\ref{eq:after1Mellin}) becomes
 \begin{eqnarray}
  \int _0 ^\infty du u ^{s+2\eta -2} (u ^2+2u \cos \theta +1)^{-1+\eta /2}
  &=& 2^{1/2-\eta /2} (\sin \theta )^{\eta /2-1/2} \frac{\Gamma (s+2\eta -1) \Gamma (3-3\eta -s)}{\Gamma (2-\eta )} \nonumber \\
   && \times \left( \frac{1+\cos \theta }{1-\cos \theta } \right) ^{\eta /4-1/4} \nonumber \\
   && \times {}_2 F_1 \left( -s-\frac{5\eta }{2}+\frac{5}{2} ,s+\frac{5\eta }{2}-\frac{3}{2},-\frac{\eta }{2}+\frac{3}{2},\frac{1-\cos \theta }{2} \right) \nonumber \\
   &\equiv & F'(s,\eta ,\theta ),\ \ \ \ \ (1-2\eta <{\rm Re}[s]<3-3\eta ) .
 \end{eqnarray}
Therefore, Eq.~(\ref{eq:polarT}) can be written as  
 \begin{eqnarray}
  T_0(\hat{\k},s) &=& \frac{1}{8\pi ^3} \int d\Omega F(s,\eta ,\phi )
  F'(s,\eta ,\theta ), \ \ \ \ \ (1-2\eta <{\rm Re}[s]<2-\eta ).
  \label{eq:MellinT}
 \end{eqnarray} 
Performing the inverse Mellin transformation, we obtain
 \begin{eqnarray}
  T_0({\hat\k},\v) &=& \frac{1}{2\pi i} 
\int _{c-i\infty } ^{c+i\infty } ds T_0(\hat{\k},s) v ^{-s} \nonumber \\
   &=& \frac{1}{2\pi i} \int _{c-i\infty } ^{c+i\infty } ds \frac{1}{8\pi ^3} \int d\Omega F(s,\eta ,\phi ) F'(s,\eta ,\theta ) v ^{-s} ,
   \label{invMellin}
 \end{eqnarray} 
where $1-2\eta <c<2-\eta $.
For the integration over $s$, we close the integration path in 
the right semicircle. Then, the poles of $F(s,\eta ,\phi )$ and 
$F'(s,\eta ,\theta )$ are, respectively, at 
 \begin{align}
  s=& 2-\eta , \ 3-\eta ,\ \cdots , \\
  s=& 3-3\eta , \ 4-3\eta ,\ \cdots .
 \end{align}
Poles we are now interested in are 
$s=2-\eta$ for diagram (c), and  $s=3-3\eta$ for diagram (e) 
which yield the $\ln k$ contributions 
(see Eqs.~(\ref{residue1}), (\ref{residue3})).
Below we evaluate $T_0(\hat\k, \v)$ only at these poles. 

First consider the pole $s=2-\eta$. 
Since the residue of the gamma function at $z=-n$ is
 \begin{eqnarray}
 \lim_{z \to -n} (z+n) \Gamma (z)  = \frac{(-1)^n}{n!},
 \end{eqnarray}
Residues of $F$ and $F'$ at $s=2-\eta$ can be evaluated as 
 \begin{eqnarray}
 \lim_{s \to 2-\eta} (s-2+\eta)  F(s ,\eta ,\phi )
 &=& 2^{1/2-\eta /2} \, \frac{1}{2} \left( \cos
  \frac{\phi }{2} \right) ^{-1-\eta } (1+\cos \phi )^{1/2+\eta /2} =1
  ,\\ 
 \lim_{s \to 2-\eta} (s-2+\eta)  F'(s ,\eta ,\theta )
 &=& 2^{1/2-\eta /2} (1+\cos \theta  )^{\eta /2-1/2} 
  \frac{\Gamma (1+\eta )\Gamma (1-2\eta )}{\Gamma
    (2-\eta )} 
 \nonumber \\ 
    && \times  {}_2 F_1 
    \left( \frac{1-3\eta }{2},\frac{1+3\eta }{2},
           \frac{3-\eta  }{2},\frac{1-\cos \theta }{2} 
    \right) .
 \end{eqnarray}  
Substituting these into Eq.~(\ref{invMellin}), we get
 \begin{eqnarray}
  \frac{1}{8\pi ^3} \int d\Omega F(s,\eta ,\phi ) F'(s,\eta ,\theta ) v ^{-2+\eta }
   &=& \frac{1}{4\pi ^2} \int _{-1} ^1 d\cos \theta 2^{1/2-\eta /2} (1+\cos \theta )^{\eta /2-1/2} 
    \frac{\Gamma (1+\eta )\Gamma (1-2\eta )}{\Gamma (2-\eta )} \nonumber \\
    && \times  {}_2 F_1 \left( \frac{1-3\eta }{2},\frac{1+3\eta }{2},\frac{3-\eta }{2},\frac{1-\cos \theta }{2} \right) v ^{-2+\eta } \nonumber \\
    &\equiv & L_1 v ^{-2+\eta } .
 \end{eqnarray}

Next, consider the other pole at $s=3-3\eta $. Then, 
 \begin{eqnarray}
 \lim_{s \to 2-\eta} (s-3+3\eta)  F(s ,\eta ,\phi ) 
&=&  2^{1/2-\eta /2} \, 
     (1+\cos \phi )^{\eta /2-1/2} 
    \frac{\Gamma (3-3\eta )\Gamma (-1+2\eta )}
         {\Gamma (2-\eta )} 
\nonumber \\
&&\times {}_2 F_1 \left( \frac{5-5\eta }{2},\frac{-3+5\eta }{2},
                         \frac{3-\eta }{2},\frac{1-\cos \phi }{2} \right),\\
 \lim_{s \to 2-\eta} (s-3+3\eta)   F'(s ,\eta ,\theta ) &=& 1 \, .
 \end{eqnarray}
Therefore, Eq.~(\ref{invMellin}) becomes
 \begin{eqnarray}
  \frac{1}{8\pi ^3} \int d\Omega F(s,\eta ,\phi ) F'(s,\eta ,\theta ) v ^{-3+3\eta }
   &=& \frac{1}{4\pi ^2} \int _{-1} ^1 d\cos \phi 2^{1/2-\eta /2} (1+\cos \phi )^{\eta /2-1/2} 
    \frac{\Gamma (3-3\eta )\Gamma (-1+2\eta )}{\Gamma (2-\eta )} \nonumber \\
    && \times  {}_2 F_1 \left( \frac{5-5\eta }{2},\frac{-3+5\eta }{2},\frac{3-\eta }{2},\frac{1-\cos \phi }{2} \right) v ^{-3+3\eta } \nonumber \\
   &\equiv & L_2 v ^{-3+3\eta } .
 \end{eqnarray}
As a result, $T_0$ becomes
 \begin{eqnarray}
  T_0({\hat\k},\v) =L_1 v ^{-2+\eta }+L_2 v ^{-3+3\eta } 
+(\mbox{higer order}) ,
 \end{eqnarray}  
and the contribution of diagram (c) is 
 \begin{eqnarray}
  {\rm (c)} &\sim & \frac{6}{\lambda {\cal A}(\eta )} k^{1-2\eta } 
\left[\frac{4}{N} \int \frac{d^3v}{8\pi ^3} 
      v ^{1-2\eta } \vert\v+{\hat\k}\vert ^{1-2\eta }\, 
      \vert \v +{\hat\k}/2 \vert ^{-2+\eta }\,  T_0({\hat\k},\v) 
  \right]\nonumber \\
      &\sim& \frac{6}{\lambda {\cal A}(\eta )} k^{1-2\eta } 
\left[\frac{4}{N} \int^{\Lambda/k} \frac{d^3v}{8\pi ^3} 
      v ^{1-2\eta } \vert \v+{\hat\k}\vert^{1-2\eta } 
      \vert \v +{\hat\k}/2 \vert ^{-2+\eta } 
        L_2 v ^{-3+3\eta }\right] \nonumber \\
      &\sim & \frac{6}{\lambda {\cal A}(\eta )} k^{1-2\eta } 
\left[\frac{4L_2}{N} \frac{1}{2\pi ^2} \ln \frac{\Lambda }{k}\right]\, .
 \end{eqnarray}  
Similarly, diagram (e) is
 \begin{eqnarray}
  {\rm (e)} &\sim & \frac{6}{\lambda {\cal A}(\eta )} k^{1-2\eta } 
\left[ \frac{-4}{N{\cal A}(\eta )} \int \frac{d^3v}{8\pi ^3} 
      v ^{1-2\eta } \vert \v+{\hat\k}\vert^{1-2\eta } \, T^2_0 ({\hat\k},\v) 
\right] \nonumber \\
      &\sim& \frac{6}{\lambda {\cal A}(\eta )} k^{1-2\eta }  
\left[\frac{-4}{N{\cal A}(\eta )} \int \frac{d^3v}{8\pi ^3} 
      v ^{1-2\eta } \vert\v+{\hat\k}\vert^{1-2\eta } 
        L_1L_2 v ^{-5+4\eta }\right] \nonumber \\
      &=& \frac{6}{\lambda {\cal A}(\eta )} k^{1-2\eta } 
\left[ \frac{-4L_1L_2}{N{\cal A}(\eta )} \frac{1}{2\pi ^2} \ln \frac{\Lambda }{k}\right] .  
 \end{eqnarray}
Notice that both diagrams (c) and (e) have the similar structure 
as those of diagrams (b) and (d). What remains is the estimation 
of the coefficients $L_1$ and $L_2$. 
We evaluated $L_1$ and $L_2$ numerically, and found that 
$L_1\sim {\cal O}(1)$ while $L_2$ is consistent with zero. 
Therefore, we conclude that diagrams (c) and (e) could have the 
$\ln k$ contributions, but are numerically very small, and can 
be ignored in our calculation.

\end{document}